\documentclass[aps,prd,amsmath,twocolumn,amssymbaps,showpacs]{revtex4-1}
\usepackage{graphicx}  
\usepackage{dcolumn}   
\usepackage{bm}        
\usepackage{amssymb}   
\usepackage{amsmath}   
\usepackage{lineno}
\usepackage{draftcopy}
\usepackage{relsize} 
\usepackage{xfrac}    
\usepackage{slashed}

\usepackage{color}
\definecolor{dgreen}{cmyk}{1.,0.,1.,0.2}        
\definecolor{orange}{cmyk}{0.,0.353,1.,0.}    

\usepackage[bookmarks]{hyperref}

\newcommand{\be}{\begin{equation}}
\newcommand{\ee}{\end{equation}}                                                                               
\newcommand{\bea}{\begin{eqnarray}}
\newcommand{\eea}{\end{eqnarray}}

\begin{document}
\title{A field theoretical model for quarkyonic matter}

\author{Gaoqing Cao}\email{caogaoqing@mail.sysu.edu.cn }
\affiliation{School of Physics and Astronomy, Sun Yat-sen University, Guangzhou 510275, China.} 
\author{Jinfeng Liao}\email{liaoji@indiana.edu}
\affiliation{Physics Department and Center for Exploration of Energy and Matter,
Indiana University, 2401 N Milo B. Sampson Lane, Bloomington, Indiana 47408, USA.}

\date{\today}
\begin{abstract} 
The possibility that nuclear matter at a density relevant to the interior of massive neutron stars may be a quarkynoic matter has  attracted considerable recent interest. In this work, we construct a  field theoretical model to describe the quarkyonic matter, that would allow  quantitative and systematic calculations of its various properties. This is implemented by synthesizing the  Walecka model  together with the quark-meson model, where both quark and nucleon degrees of freedom are present based on the  quarkyonic scenario. With this model we compute at mean-field level the  thermodynamic properties of the symmetric nuclear matter   and calibrate model parameters through well-known nuclear physics measurements. We find this model gives a very good description of the symmetric nuclear matter from moderate to high baryon density and demonstrates a continuous transition from nucleon-dominance to quark-dominance for the system.   
\end{abstract}

\pacs{11.30.Qc, 05.30.Fk, 11.30.Hv, 12.20.Ds}

\maketitle

\section{Introduction}\label{introduction}

To understand the phases and properties of strong interaction matter at high baryon density, especially in the region relevant to the interior of massive neutron stars, is a very active frontier in the research field of nuclear physics and nuclear astrophysics. The study of high baryon density region is also very relevant to ongoing experimental measurements (e.g. STAR at RHIC and HADES at SPS) of heavy ion collisions at low beam energy as well as planned programs at future facilities like the FAIR, NICA and HIAF. 
Both neutron star observations and heavy ion experiments will help promote our understanding of the phase diagram over a broad range of temperature and baryon density for  the strong interaction matter governed by Quantum Chromodynamics (QCD). For recent reviews, see e.g.~\cite{Page:2006ud,Lattimer:2015nhk,Braun-Munzinger:2015hba,Bzdak:2019pkr,Luo:2017faz}.

While a lot has been learned about the QCD matter properties at zero or very small baryon density, the high density region remains a significant challenge.  There are several interesting proposals about possible phases of high density QCD matter, such as  (two-flavor) color superconductivity~\cite{Alford:1997zt,Rapp:1997zu}, color-flavor locking~\cite{Alford:1998mk}, or quarkyonic matter~\cite{McLerran:2007qj}. 
In the density region comparable with the neutron star interior, the quarkyonic matter might be more directly relevant, thus we shall focus on that phase in this work. The quarkyonic matter was first proposed by following insights from the large $N_c$ analysis and emphasizing the coexistence of nucleon/quark degrees of freedom~\cite{McLerran:2007qj,McLerran:2008ua,Andronic:2009gj,Kojo:2009ha}. In the large isospin density (but small baryon density) region, an analogous "quarksonic matter" was proposed by following similar arguments in Ref.~\cite{Cao:2016ats}. Recently, there has been increasing interest to explore the possible existence of quarkyonic matter inside compact stars and the consequences for relevant astrophysical observations~\cite{Steinheimer:2011ea,McLerran:2018hbz,Fukushima:2015bda,Jeong:2019lhv,Sen:2020peq,Duarte:2020xsp,Zhao:2020dvu,Xia:2018cpy}.

Certain issues require improvements over previous studies, many of which were based on simple (and often oversimplified) picture implementations with crude and ad hoc approximations. The chiral symmetry restoration with increasing density often lacked a dynamical treatment. The important physics constraints from the lower density side, e.g. nuclear matter saturation properties, were not carefully checked. Given these issues, it is therefore important to develop a  more sophisticated field theoretical model to study the quarkyonic matter -- one that would allow systematic calculations of various properties and quantitative scrutiny of important physics constraints. This is the main goal of our study, with the first successful step to be reported in the present paper. The rest of paper is organized as follows. In Sec.\ref{Formulism}, the model Lagrangian density is constructed and the main formalism is developed in great details by focusing on symmetric nuclear matter. Then in Sec.\ref{MP}, the model parameters are fixed according to the saturation properties observed from low energy nuclear experiments. In Sec.\ref{properties}, the thermodynamic properties are computed for quarkyonic matter for a wide range of baryon densities within our new model. Finally, we conclude in Sec.\ref{conclusions}.

\section{An effective model for quarkyonic matter}\label{Formulism}

\subsection{Lagrangian and thermodynamic potential}
By following the spirit of quark-baryonic (or quarkyonic) matter (QBM) with both quarks and baryons as the effective degrees of freedom of the strong interaction system, we construct a field theoretical   model which combines  the quark-meson (QM) model~\cite{Schaefer:2006ds} together with the well-known Walecka model~\cite{Walecka}. The quark-meson and Walecka models are common on one aspect: mesons are the ``messengers'' of the interactions between quarks or baryons. The overall Lagrangian density of our two-flavor model is composed of three parts, that is, ${\cal L}_{QBM}\equiv{\cal L}_{\rm q}+{\cal L}_{\rm N}+{\cal L}_{\rm M}$ with the following explicit forms:
\begin{widetext}
\begin{eqnarray}\label{QH}
{\cal L}_{\rm q}\!\!&=&\!\!\bar{q}\Big[i\slashed{\partial}+\left(\mu_B/N_c+{\mu_I\over 2}\tau_3\right)\gamma^0-g_{q}\left(\sigma+i\gamma^5\boldsymbol{\tau\cdot\pi}\right)\Big]q,\nonumber\\
{\cal L}_{\rm N}\!\!&=&\!\!\bar{N}\Big[i\slashed{\partial}+\left(\mu_B+{\mu_I\over 2}\tau_3\right)\gamma^0-g_{\rm Ns}\left(\sigma+i\gamma^5\boldsymbol{\tau\cdot\pi}\right)+g_{\rm Nv}\left(\slashed{\rho}-\gamma^5\slashed{A}\right)\boldsymbol{\cdot}\tau\Big]N,\nonumber\\
{\cal L}_{\rm M}\!\!&=&\!\!{1\over2}\left(\partial_\mu\sigma\partial^\mu\sigma+D_\mu\boldsymbol{\pi}\boldsymbol{\cdot}D^\mu\boldsymbol{\pi}\right)-{\lambda\over4}\left(\sigma^2+\boldsymbol{\pi\cdot\pi}-\upsilon^2\right)^2+c~\sigma+{1\over2}g_{\rm sv}\left(\sigma^2+\boldsymbol{\pi\cdot\pi}\right)\left(\rho_\mu\boldsymbol{\cdot}\rho^\mu+A_\mu\boldsymbol{\cdot}A^\mu\right)\nonumber\\
&&-{1\over4}\left(D_\mu\rho_\nu-D_\nu\rho_\mu\right)\boldsymbol{\cdot}\left(D^\mu\rho^\nu-D^\nu\rho^\mu\right)+{1\over2}m_v^2\rho_\mu\boldsymbol{\cdot}\rho^\mu-{1\over4}\left(D_\mu A_\nu-D_\nu A_\mu\right)\boldsymbol{\cdot}\left(D^\mu A^\nu-D^\nu A^\mu\right)+{1\over2}m_a^2 A_\mu\boldsymbol{\cdot} A^\mu \ . \quad
\end{eqnarray}
\end{widetext}
Here, the quantum fields are defined as the following: $q(x)=(u(x),d(x))^T$ denotes the two-flavor quark field with color degrees of freedom $N_c=3$, $N(x)=(p(x),n(x))^T$ is the two-flavor nucleon field {\it outside} the Fermi spheres of quarks if exist, $\sigma(x)$ and $\boldsymbol{\pi}(x)$ are   the scalar and pseudoscalar mesons, while $\rho^a_\mu$ (with $\rho^0_\mu$ the $\omega$ meson) and $A^a_\mu\ (a=0,\dots,3)$ are vector and axial vector mesons, respectively. The baryon and isospin chemical potentials are given by $\mu_B$ and ${\mu_I\over 2}$, respectively. The isospin matrices are   
$$\tau=\left(1,{\tau_x-i\tau_y\over\sqrt{2}},{\tau_x+i\tau_y\over\sqrt{2}},\tau_z\right)$$
with $\tau_{x},\tau_{y}$ and $\tau_{z}$ the Pauli matrices in flavor space. The  derivative operators are defined as $D_0=\partial_0\mp i{\mu_I}$ for the charged $\pi^\pm,\rho^\pm_\mu$ and $A_\mu^
\pm$, and $D_\mu=\partial_\mu$ for the others. For the isospin symmetric case with ${\mu_I}=0$, the Lagrangian has exact chiral symmetry in the chiral limit $c=0$ and when chiral anomaly is neglected by choosing $m_v=m_a$. In the realistic case, with the linear coefficient $c\neq0$ and the masses $m_v<m_a$, there is only approximate chiral symmetry in the QBM model. 

Let us first discuss the vacuum of the above model at   temperature $T=0$ and chemical potential $\mu=0$. 
In mean field approximation, the thermodynamic potential is only given by the mesonic part in the vacuum, that is,
\begin{eqnarray}
\Omega_v={\lambda\over4}\left(\langle\sigma\rangle^2+{\langle\boldsymbol{\pi}\rangle\cdot\langle\boldsymbol{\pi}\rangle}-\upsilon^2\right)^2-c~\langle\sigma\rangle,
\end{eqnarray}
the global minimum of which locates at $\langle\boldsymbol{\pi}\rangle=0$ and 
\begin{eqnarray}\label{sigma0}
\langle\sigma\rangle=\sum_{t=\pm}\left[{{c\over2\lambda}+t~i\sqrt{{\upsilon^6\over27}-\Big({c\over{2\lambda}}\Big)^2}}\right]^{1/3}.
\end{eqnarray}
It can be checked that we simply have $\langle\sigma\rangle=\upsilon$ in the chiral limit $c\rightarrow0$. Based on the ground state, the sigma and pion masses can then be derived as~\cite{Schaefer:2006ds}
\begin{eqnarray}
m_\sigma^2=\lambda\left(3\langle\sigma\rangle^2-\upsilon^2\right),~m_\pi^2=\lambda\left(\langle\sigma\rangle^2-\upsilon^2\right),
\end{eqnarray}
which indicate the $\sigma$ and $\pi$ mesons as the massive and Goldstone modes, respectively.

If we adopt the quark version of Goldberger-Treiman relation: $f_\pi^2g_{q}^2=m_{\rm q}^2$~\cite{Klevansky:1992qe}, the expectation value of $\sigma$ is found to be $\langle\sigma\rangle_v= f_\pi$ in vacuum. Then, the parameters in the mesonic sector can be determined by the vacuum masses $m_\sigma,m_\pi$ and pion decay constant $f_\pi$ as
\begin{eqnarray}\label{mpara}
\lambda={m_\sigma^2-m_\pi^2\over2f_\pi^2},~\upsilon^2={m_\sigma^2-3m_\pi^2\over m_\sigma^2-m_\pi^2}f_\pi^2,~c=f_\pi m_\pi^2.
\end{eqnarray}
We next discuss the other model parameters in the quark and baryonic sectors. Firstly, the coupling constants between the scalar sector mesons and quarks or nucleons can be fixed by their vacuum masses as $g_{\rm q}=m_{\rm q}^v/f_\pi\equiv m_\sigma/(2f_\pi)$~\cite{Schaefer:2006ds} and  $g_{\rm Ns}=m_{\rm N}^v/f_\pi$. The quantities $m_\pi, f_\pi$ and $m_{\rm N}^v$ are well determined from the experiments. The other parameters like  $m_{\rm q}^v$ (or $m_\sigma$), $g_{\rm Nv}$ and $g_{\rm sv}$ will be  constrained later by  the empirical saturation properties of nuclear matter. Note also that with the additional scalar-vector interaction, the vector mass is given by $m_{\rm v}^2+g_{\rm sv}f_\pi^2=(785~{\rm MeV})^2$ in the vacuum. 

We now turn to compute thermodynamics at finite temperature and chemical potentials, where quarks and nucleons will also give  contributions. In this paper, we will focus on the (isospin-)symmetric nuclear matter  as a first step, by choosing $\mu_B>0$ and $\mu_I=0$. The thermodynamic contributions from the quark and baryon sectors are given below: 
\begin{widetext}
\begin{eqnarray}
\Omega_{q}^t&=&-4N_cT\!\!\sum_{t=\pm}\!\int{d^3p\over(2\pi)^3}~\ln\left(1+e^{-\left[E_{\rm q}({\bf p})+t{\mu_B\over N_c}\right]/T}\right),\\
\Omega_{N}^t&=&-{1\over2}\left(g_{\rm sv}\langle\sigma\rangle^2+m_{\rm v}^2\right)\left(\langle\omega_0\rangle^2+(\langle\rho_0^3\rangle)^2\right)-4T\sum_{t=\pm}\int{d^3p\over(2\pi)^3}~\ln\left({1+e^{-[E_{\rm N}({\bf p})+t(\mu_B-g_{\rm Nv} \langle\omega_0\rangle)]/T}\over 1+e^{-[E_{\rm N}({\bf p})+t({\mu_{B}'}-g_{\rm Nv} \langle\omega_0\rangle)]/T}}\right), \ \quad
\end{eqnarray}
\end{widetext}
where the dispersion relations are $E_{\rm q}({\bf p})=\left({\bf p}^2+m_{\rm q}^2\right)^{1/2}$ with $m_{\rm q}=g_{\rm q}\langle\sigma\rangle$ and $E_{\rm N}({\bf p})=\left({\bf p}^2+m_{\rm N}^2\right)^{1/2}$ with $m_{\rm N}=g_{\rm Ns}\langle\sigma\rangle$. The vector mean-field condensate is subject to the physical constraint $0\leq g_{\rm Nv} \langle\omega_0\rangle\leq\mu_B$, that is, the nucleon chemical potential  is reduced by $\langle\omega_0\rangle$ but never changes sign.

The  crucial step here is to implement the quarkyonic picture in the momentum space, in which the interior of the Fermi sea is filled up by quarks while the nucleons are excluded to reside in an outside shell of the Fermi sea~\cite{McLerran:2007qj,McLerran:2008ua}.  
In our model, the  boundary for  "Pauli-blocked" nucleon sphere is characterized by an effective chemical potential ${\mu_{B}'}$. The nucleons in the quarkyonic matter exist between the Fermi sphere stretched by ${\mu_{B}'}$ and $\mu_B$. As one can tell in $\Omega_{N}^t$: the thermodynamics potential of the nucleonic part is obtained by subtracting the supposed inner contribution (with ${\mu_{B}'}$) out of the naive total one (with $\mu_B$). It is important to have an appropriate scheme for determining the ${\mu_{B}'}$. One possible choice is the $\mu_B$-linear form:
\begin{eqnarray}\label{muPB1}
\mu_{B}'=\mu_B-(N_cm_{\rm q}-m_{\rm N}),
\end{eqnarray} 
based on comparing kinetic energy of a baryon with that of $N_c$ quarks. Another nonlinear choice assumes that the momenta of the valence quarks of proton ($uud$) and neutron ($udd$) are the same and nucleons are blocked by the free quarks from the Fermi sphere~\cite{McLerran:2018hbz}, that is,
\begin{eqnarray}\label{muPB2}
{\mu_{B}'}= \sqrt{m_{\rm N}^2+\left(N_ck_F\right)^2},
\end{eqnarray}
which is smaller than $\mu_B$ as $N_cm_{\rm q}>m_{\rm N}$.
Here, the effective Fermi momentum of the  $u$ and $d$ quarks is
\begin{eqnarray}
k_F=\left[\left({\mu_B/ N_c}\right)^2-m_{\rm q}^2\right]^{1/2} \ .
\end{eqnarray}
This definition  is based on comparing momentum of a baryon with that of $N_c$ quarks. We will perform computations with both choices of ${\mu_{B}'}$ and compare their results later. 

\subsection{Gap equations and energy density}
In mean field approximation, the total thermodynamic potential is then $\Omega=\Omega_v+\Omega_{q}^t+\Omega_{N}^t$ and the gap equations can be obtained from the extremal conditions $\partial\Omega/\partial X=0\ \ (X=\langle\omega_0\rangle, \langle\sigma\rangle)$ as
\begin{widetext}
\begin{eqnarray}
&&\langle\omega_0\rangle=-{4}\sum_{t=\pm}\int{d^3p\over(2\pi)^3}{t{g_{\rm Nv}\over m_{\rm v}^2+g_{\rm sv}\langle\sigma\rangle^2}\over 1+e^{[E_{\rm N}({\bf p})+t(\mu_B-g_{\rm Nv} \langle\omega_0\rangle)]/T}}+{4}\sum_{t=\pm}\int{d^3p\over(2\pi)^3}{t{g_{\rm Nv}\over m_{\rm v}^2+g_{\rm sv}\langle\sigma\rangle^2}\over 1+e^{[E_{\rm N}({\bf p})+t({\mu_{B}'}-g_{\rm Nv} \langle\omega_0\rangle)]/T}},\label{omega0}\\
&&{\lambda}\left(\langle\sigma\rangle^2-\upsilon^2\right)\langle\sigma\rangle-c-g_{\rm sv}\langle\sigma\rangle\langle\omega_0\rangle^2+4N_c\sum_{t=\pm}\int{d^3p\over(2\pi)^3}{{g_{q}m_{\rm q}/ E_{\rm q}({\bf p})}\over1+e^{[E_{\rm q}({\bf p})+t{\mu_B\over N_c}]/T}}+4\sum_{t=\pm}\int{d^3p\over(2\pi)^3}{{g_{\rm Ns}m_{\rm N}/E_{\rm N}({\bf p})}\over 1+e^{[E_{\rm N}({\bf p})+t(\mu_B-g_{\rm Nv} \langle\omega_0\rangle)]/T}}\nonumber\\
&&-4\sum_{t=\pm}\int{d^3p\over(2\pi)^3}{{g_{\rm Ns}m_{\rm N}/E_{\rm N}({\bf p})}+t\,{\partial{\mu_{B}'}/\partial\langle\sigma\rangle}\over 1+e^{[E_{\rm N}({\bf p})+t({\mu_{B}'}-g_{\rm Nv} \langle\omega_0\rangle)]/T}}=0,\label{sigma}
\end{eqnarray}
where the derivatives of the effective chemical potential are    
${\partial{\mu_{B}'}\over \partial\langle\sigma\rangle}=g_{\rm Ns}-g_{\rm q}N_{\rm c}$ for the linear choice and $ {\partial{\mu_{B}'}\over \partial\langle\sigma\rangle}={1\over {\mu_{B}'}}\left[g_{\rm Ns}m_{\rm N}-g_{\rm q}N_{\rm c}^2m_{\rm q}\right]$
for the  nonlinear choice, respectively.

Furthermore, the baryon number and entropy densities can be derived directly according to the thermodynamic relationships $n_B=-\partial\Omega/\partial\mu_B$ and $s=-\partial\Omega/\partial T$ as:
\begin{eqnarray}
n_B&=&-4\sum_{t=\pm}\int{d^3p\over(2\pi)^3}t\left({1\over 1+e^{[E_{\rm q}({\bf p})+t{\mu_B\over N_c}]/T}}+{1\over 1+e^{[E_{\rm N}({\bf p})+t(\mu_B-g_{\rm Nv} \langle\omega_0\rangle)]/T}}-{{\partial{\mu_{B}'}/ \partial\mu_B}\over 1+e^{[E_{\rm N}({\bf p})+t({\mu_{B}'}-g_{\rm Nv} \langle\omega_0\rangle)]/T}}\right),\label{Bnumber}\\
s&=&4\sum_{t=\pm}\int{d^3p\over(2\pi)^3}\left(N_c\ln\Big(1+e^{-[E_{\rm q}({\bf p})+t{\mu_B\over N_c}]/T}\Big)+{N_cE_{\rm q}({\bf p})\!+\!t\,{\mu_B}\over T\left(1\!+\!e^{[E_{\rm q}({\bf p})+t{\mu_B\over N_c}]/T}\right)}+\ln\Big(1+e^{-[E_{\rm N}({\bf p})+t(\mu_B-g_{\rm Nv} \langle\omega_0\rangle)]/T}\Big)\right.+\nonumber\\
&&\left.{E_{\rm N}({\bf p})\!+\!t(\mu_B-g_{\rm Nv} \langle\omega_0\rangle)\over T\left(1\!+\!e^{[E_{\rm N}({\bf p})+t(\mu_B-g_{\rm Nv} \langle\omega_0\rangle)]/T}\right)}-\ln\Big(1\!+\!e^{-[E_{\rm N}({\bf p})+t({\mu_{B}'}-g_{\rm Nv} \langle\omega_0\rangle)]/T}\Big)-{E_{\rm N}({\bf p})\!+\!t({\mu_{B}'}\!-\!g_{\rm Nv} \langle\omega_0\rangle)\over T\left(1\!+\!e^{[E_{\rm N}({\bf p})+t({\mu_{B}'}-g_{\rm Nv} \langle\omega_0\rangle)]/T}\right)}\right),
\end{eqnarray}
where the explicit forms of the derivatives of the effective chemical potentials in Eq.\eqref{Bnumber} are given by 
${\partial{\mu_{B}'}\over \partial\mu_B}=1$ for the linear choice and   
${\partial{\mu_{B}'}\over \partial\mu_B}={\mu_B\over {\mu_{B}'}}$ 
for  the nonlinear choice, respectively. 
Thus, the   energy density of the quarkyonic matter is found to be 
\begin{eqnarray}
\epsilon&\equiv&\Omega+\mu_Bn_B+sT-(T=\mu_B=0)\nonumber\\
&=&{\lambda\over4}\Big(\langle\sigma\rangle^2\!-\!\upsilon^2\Big)^2\!-\!c~\langle\sigma\rangle-{1\over2}(m_{\rm v}^2\!+\!g_{\rm sv}\langle\sigma\rangle^2)\langle\omega_0\rangle^2\!+\!4\sum_{t=\pm}\int{d^3p\over(2\pi)^3}\left({N_cE_{\rm q}({\bf p})\over1\!+\!e^{[E_{\rm q}({\bf p})+t{\mu_B\over N_c}]/T}}\!+\!{E_{\rm N}({\bf p})\!-\!t\,g_{\rm Nv} \langle\omega_0\rangle\over1\!+\!e^{[E_{\rm N}({\bf p})+t(\mu_B-g_{\rm Nv} \langle\omega_0\rangle)]/T}}\right.\nonumber\\
&&\left.-{E_{\rm N}({\bf p})\!+\!t\left[({\mu_{B}'}-\mu_B{\partial{\mu_{B}'}/ \partial\mu_B})-g_{\rm Nv} \langle\omega_0\rangle\right]\over 1+e^{[E_{\rm N}({\bf p})+t({\mu_{B}'}-g_{\rm Nv} \langle\omega_0\rangle)]/T}}\right)-(T=\mu_B=0),\label{energy}
\end{eqnarray}
where we assume $m_{\rm N}^2+(N_{\rm c}k_F)^2>0$ and the vacuum term is excluded to make sure the medium energy vanishes in the vacuum.

To fix the remaining parameters of the model, we turn to zero temperature limit where some empirical results are well known. The explicit form of the baryon density Eq.\eqref{Bnumber} for the quarkyonic matter becomes
\begin{eqnarray}
n_B&=&n_B^q+n_B^N-n_B^{N'}\equiv {2p_{\rm qF}^3\over3\pi^2}+{2p_{\rm NF}^3\over3\pi^2}-{\partial{\mu_{B}'}\over \partial\mu_B}{2p_{\rm N'F}^3\over3\pi^2},\label{density}
\end{eqnarray}
where $p_{\rm NF}$ and $p_{\rm N'F}$ are the Fermi momenta of the occupied  and Pauli-blocked nucleon states, and $p_{\rm qF}$ is the Fermi momentum of the occupied quark states, respectively. The Fermi momenta are related to the chemical potentials through the Fermi energies as
\begin{eqnarray}
E_{\rm qF}\equiv E_{\rm q}(p_{\rm qF})=\mu_B/N_c,\ E_{\rm NF}\equiv E_{\rm N}(p_{\rm NF})=\mu_B-g_{\rm Nv} \langle\omega_0\rangle,\ 
E_{\rm N'F}\equiv E_{\rm N}(p_{\rm N'F})={\mu_{B}'}-g_{\rm Nv} \langle\omega_0\rangle.
\end{eqnarray}
In this case, we're glad that the momentum integrations involved in the gap equations Eqs.\eqref{omega0} and \eqref{sigma}
 and energy density Eq.(\ref{energy}) can be carried out explicitely with the help of Fermi momenta as
\begin{eqnarray}
0&=&\langle\omega_0\rangle(m_{\rm v}^2+g_{\rm sv}\langle\sigma\rangle^2)-g_{\rm Nv}~\Delta{2p_{\rm NF}^3\over3\pi^2},\label{omega01}\\
0&=&{\lambda}\left(\langle\sigma\rangle^2-\upsilon^2\right)\langle\sigma\rangle-c-g_{\rm sv}\langle\sigma\rangle\langle\omega_0\rangle^2+{g_{\rm Ns}m_{\rm N}\over{\pi^2}}\Delta\left[E_{\rm NF}p_{\rm NF}-m_{\rm N}^2\ln\Big({E_{\rm NF}+p_{\rm NF}\over m_{\rm N}}\Big)\right]\nonumber\\
&&+N_c{g_{q}m_{\rm q}\over{\pi^2}}\left[E_{\rm qF}p_{\rm qF}-m_{\rm q}^2\ln\Big({E_{\rm qF}+p_{\rm qF}\over m_{\rm q}}\Big)\right]+{\partial{\mu_{B}'}\over \partial\langle\sigma\rangle}{2p_{\rm N'F}^3\over3\pi^2},\\
\epsilon&=&{\lambda\over4}\left(\langle\sigma\rangle^2-\upsilon^2\right)^2-c~\langle\sigma\rangle+{1\over2}(m_{\rm v}^2+g_{\rm sv}\langle\sigma\rangle^2)\langle\omega_0\rangle^2+{1\over4\pi^2}\Delta\left[2E_{\rm NF}^3p_{\rm NF}-m_{\rm N}^2E_{\rm NF}p_{\rm NF}-m_{\rm N}^4\ln\Big({E_{\rm NF}+p_{\rm NF}\over m_{\rm N}}\Big)\right]\nonumber\\
&&+{N_c\over4\pi^2}\left[2E_{\rm qF}^3p_{\rm qF}-m_{\rm q}^2E_{\rm qF}p_{\rm qF}-m_{\rm q}^4\ln\Big({E_{\rm qF}+p_{\rm qF}\over m_{\rm q}}\Big)\right]+\left({\mu_{B}'}-\mu_B{\partial{\mu_{B}'}\over\partial\mu_B}\right){2p_{\rm N'F}^3\over3\pi^2}-(\mu_B=0),
\end{eqnarray}
\end{widetext}
where the symbol $"\Delta"$ means excluding the corresponding one with $N\rightarrow N'$ for the energy and momentum. Combining Eqs.\eqref{density} and \eqref{omega01}, we find $\langle\omega_0\rangle(m_{\rm v}^2+g_{\rm sv}\langle\sigma\rangle^2)=g_{\rm Nv}~\Delta n_B^N$ for the linear choice, which actually has a definite physical meaning: the vector condensate is proportional to the nucleon density~\cite{Walecka}.

\section{Model parameters}\label{MP}

From the experimental measurements associated with finite nuclei, some properties of the infinite and isospin symmetric nuclear matter were  extracted: the saturation density $n_0\approx0.16~{\rm fm}^{-3}$~\cite{Hofstadter}, the energy per nucleon at this density $E/N\equiv\epsilon/n_B-m_{\rm N}^v=-16~{\rm MeV}$~\cite{Green1953,Green1954} as well as the compressibility $K_0=240\pm20~{\rm MeV}$~\cite{Shlomo2006} . Theoretically, they are related with each other as:
\begin{eqnarray}
{\partial(E/N)\over\partial n_B}\Bigg|_{n_B=n_0}^{\langle\sigma\rangle}&=&-{\epsilon\over n_0^2}+{\mu_{Bc}\over n_0}={P_c\over n_0}=0,\label{saturation}\\
K_0&=&9{\partial^2(E/N)\over\partial(n_B/n_0)^2}\Bigg|_{n_B=n_0},
\end{eqnarray}
from which it is easy to infer the pressure $P_c=0$ and the critical chemical potential $\mu_{Bc}=E/N+m_{\rm N}^v=923~{\rm MeV}$. 

\begin{figure}[!htb]
	\centering
	\includegraphics[width=0.42\textwidth]{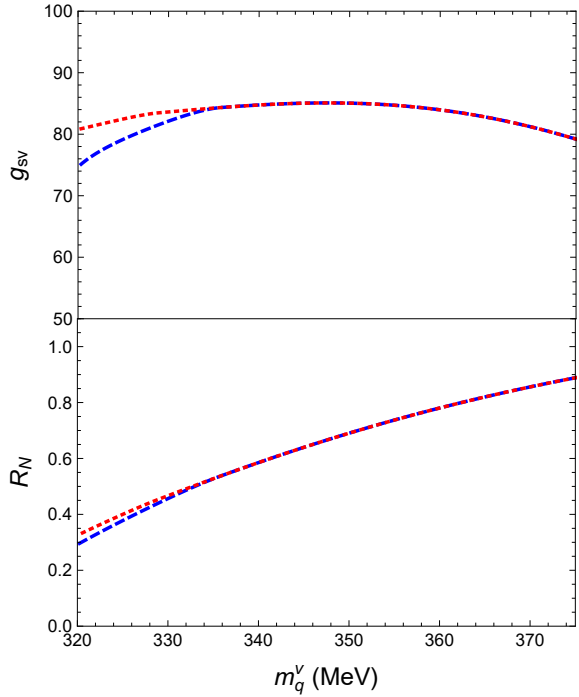}
	\caption{The extracted values for the coupling $g_{\rm sv}$ and the associated nucleon fraction $R_N$ as functions of the quark vacuum mass $m_{\rm q}^v$ for both linear (red dotted) and nonlinear (blue dashed) ${\mu_{B}'}$ choices.}\label{fitting}
\end{figure}

Now, we use these saturation properties to fix the remaining parameters.  The dynamical quark mass is varied in the range $m_{\rm q}^v\gtrsim m_{\rm N}^v/3$ which guarantees the stability of nucleons against the decay to quarks in vacuum. Then the order parameters $\langle\sigma\rangle$ and $\langle\omega_0\rangle$ and coupling constants $g_{\rm Nv}$ and $ g_{\rm sv}$ are fixed by solving the gap equations Eqs.\eqref{omega0} and \eqref{sigma}, saturation equation Eq.\eqref{saturation} and the saturation energy $E/N=-16~{\rm MeV}$ self-consistently. The extracted results for $g_{\rm sv}$ and the associated nucleon fraction $R_N\equiv\Delta n_B^N/n_B$ as functions  of $m_{\rm q}^v$ are shown together in Fig.~\ref{fitting} for both linear and nonlinear ${\mu_{B}'}$ choices. As we can see, the results are quantitatively consistent with each other for these choices, with only minor differences in the relatively smaller mass region. 
 
\begin{figure}[!htb]
	\centering
	\includegraphics[width=0.45\textwidth]{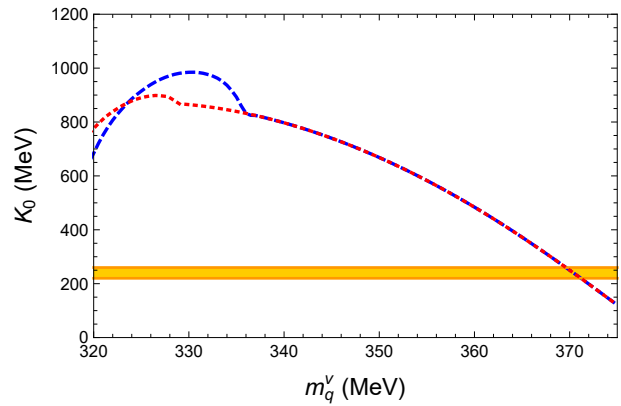}
	\caption{The compressibility $K_0$ of quarkyonic matter at saturation density $n_0$ as a function of the quark vacuum mass $m_{\rm q}^v$ for both linear (red dotted) and nonlinear (blue dashed) ${\mu_{B}'}$ choices. The yellow band is the constraint from experiments~\cite{Shlomo2006}.}\label{comp}
\end{figure}

In order to further fix the  vacuum quark mass in our model, we show our model calculations together with the empirical constraint~\cite{Shlomo2006} in Fig.\ref{comp} for the compressibility at saturation density $n_0$. From the results, we find the best agreement is achieved for $m_{\rm q}^v=370.4\pm0.8~{\rm MeV}$. In the rest of this paper, we will  then adopt the value $m_{\rm q}^v=370.4~{\rm MeV}$, with the corresponding nucleon ratio $R_N\approx85.8\%$ for both choices of ${\mu_{B}'}$ at saturation density. We note that at this density there is a nonzero albeit very small fraction of quarks that already emerge and coexist with the nucleons.   The corresponding coupling constants are also fixed to be $g_{\rm Nv}\approx7.2$ and $ g_{\rm sv}\approx81$, respectively.

\begin{figure}[!htb]
	\centering
	\includegraphics[width=0.45\textwidth]{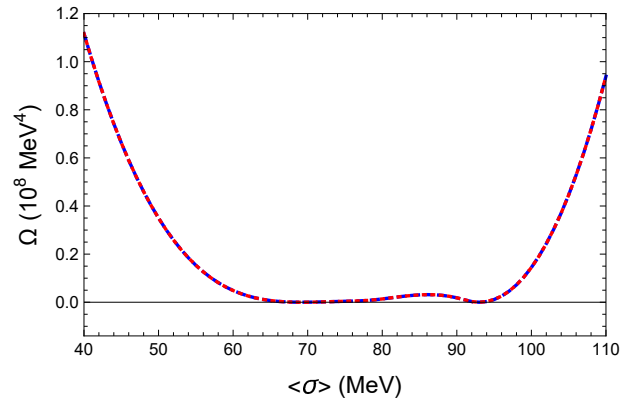}
	\caption{The thermodynamic potential $\Omega$ as a function of the chiral condensate $\langle\sigma\rangle$  at the critical chemical potential $\mu_{Bc}$. The conventions are the same as those in Fig.\ref{fitting}.}\label{Omega}
\end{figure}

At this point, all of our model parameters are fixed and the model satisfactorily catches the nuclear matter properties at saturation density. Lastly we examine the liquid-gas transition at this density. In Fig.~\ref{Omega}, we show the thermodynamic potential $\Omega$   as a function of the quark condensate $\langle\sigma\rangle$ for both choices of ${\mu_{B}'}$ at the  critical chemical potential $\mu_{Bc}$. As one can see, there is a typical first-order transition structure with two degenerate minima: one at the vacuum value $\langle\sigma\rangle=f_\pi$, and the other new one at $\langle\sigma\rangle=69.2~{\rm MeV}$. At $\mu_{Bc}$, the chiral condensate jumps from the vacuum value to the smaller one.   In the next section we will analyze the matter properties at chemical potential beyond this transition point. 
 
\section{The quarkyonic matter properties}\label{properties}

\begin{figure}[!htb]
	\centering
	\includegraphics[width=0.42\textwidth]{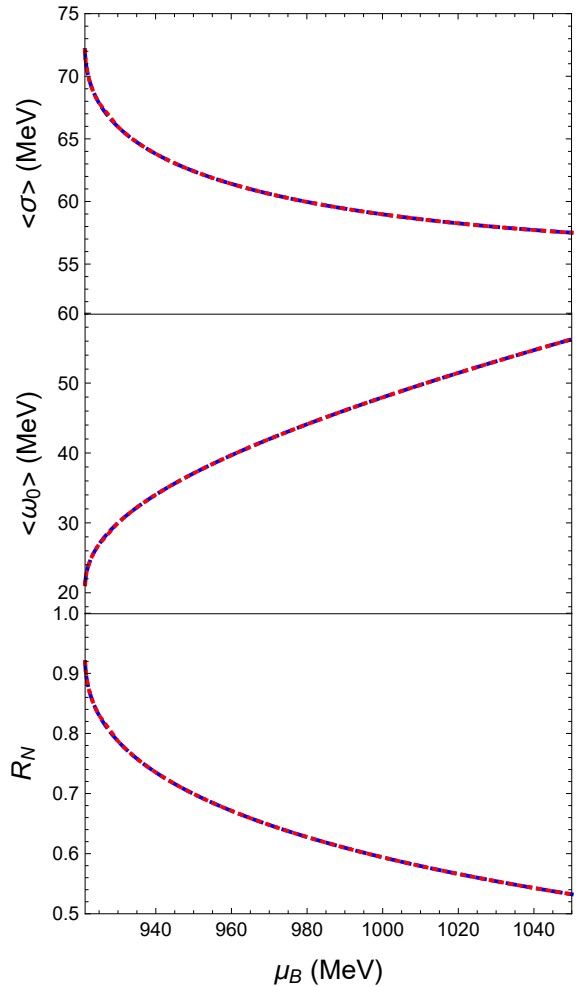}
	\caption{The chiral condensate $\langle\sigma\rangle$, vector condensate $\langle\omega_0\rangle$ and nucleon ratio $R_{\rm N}$ as functions of chemical potential $\mu_{B}$ in the chiral symmetry partially restored phase. The conventions are the same as those in Fig.\ref{fitting}.}\label{condensates}
\end{figure}

In this section we present results for quarkyonic matter properties in the region of a few times the saturation density. The chiral and vector condensates as well as the corresponding nucleon ratio are shown  in Fig.~\ref{condensates} as functions of baryon chemical potential.  We find that   both $\langle\sigma\rangle$ and $R_N$ decreas rapidly while $\langle\omega_0\rangle$ increases with $\mu_B$. This implies that with increasing density, the chiral symmetry gets gradually restored with the quarks becoming lighter and more abundant. 
The increasing of $\langle\omega_0\rangle$ could be understood as due to the enhancement of the nucleon density with $\mu_B$, even though the nucleon fraction $R_N$ decreases.

\begin{figure}[!htb]
	\centering
	\includegraphics[width=0.42\textwidth]{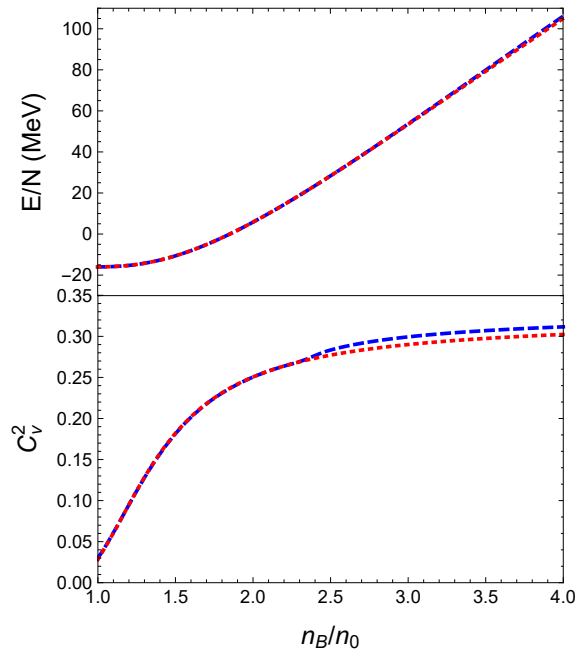}
	\caption{The energy per nucleon $E/N$ and speed of sound $C_v^2$ as functions of baryon density $n_B$, the range of which corresponds to  that of $\mu_B$ in Fig.~\ref{condensates}. The conventions are the same as those in Fig.\ref{fitting}.}\label{SV}
\end{figure}

We now compute the energy density of the system and present the closely related $E/N$ in the upper panel of Fig.~\ref{SV}.  
As we can see, the $E/N$ starts from the minimum value of $-16\,\rm MeV$  at the saturation density $n_0$ and steadily increases toward higher density. A key quantity related to the equation-of-state (EOS) for the quarkyonic matter is the speed of sound $C_v \equiv \sqrt{{\partial P\over\partial \epsilon}}$~\cite{Baym:2017whm}.  In the lower panel of Fig.~\ref{SV}, we show $C_v^2$ versus baryon density for both linear and nonlinear ${\mu_{B}'}$ choices, which show small deviation from each other. 
In both cases, the speed of sound increases quickly between $1\sim 2 \rm n_0$ and then  approaches the high density asymptotical limit rather smoothly, in consistency with a continuous transition feature~\cite{Baym:2017whm}.  We note that our results are consistent with those given in Ref.~\cite{McLerran:2018hbz,Jeong:2019lhv} for both small and large density, except that the prominent peak structure in the intermediate density is absent in our model. The difference could be due to the hard core feature in Ref.~\cite{Jeong:2019lhv} which we do not have. Actually, the monotonous feature of $C_v$ was also found in a recent quite convincing study when diquark dynamics is ignored~\cite{Leonhardt:2019fua}.

\begin{figure}[!htb]
	\centering
	\includegraphics[width=0.45\textwidth]{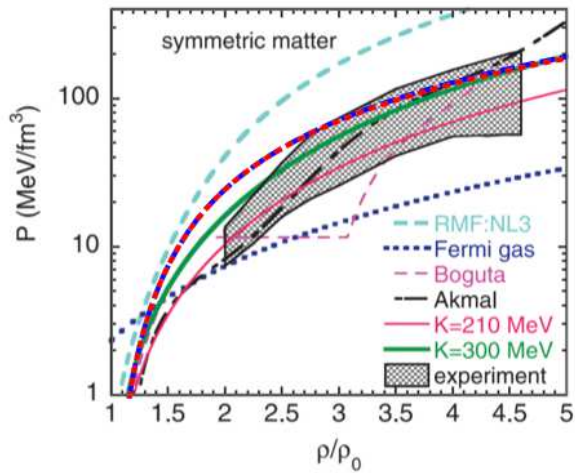}
	\caption{The comparison between our quarkyonic matter model results with those from experimental constraints (shadow region) and other model predictions (colored lines) for the pressure $P$ of symmetric nuclear matter as a function of baryon density. Note that this plot is made via adapting an original figure extracted from Re.~\cite{Danielewicz2002}, for which   we keep the original notations of the various physical quantities. In particular, the baryon density $\rho~(\rho_0)$ in this plot corresponds to $n_B~(n_0)$ we use in other places of the paper.}\label{Pressure_nB}
\end{figure}

Finally, we proceed to compare our EOS with the experimental extraction as well as other model calculations~\cite{Danielewicz2002},
see Fig.~\ref{Pressure_nB}. 
The comparison indicates that  our results based on quarkyonic matter are reasonably consistent with the experimental constraints, especially in the large density region where quarks become more and more important.

\section{conclusions}\label{conclusions}
In this work, we propose a field theoretical model for quarkyonic matter by combining the Walecka model together with the quark-meson model. 
We have systematically calibrated the model parameters based on various hadron properties in the vacuum as well as nuclear matter properties a the saturation density. Based on that, we then extend our calculations to the large baryon density region and find a number of interesting results.  Firstly, the chiral symmetry is partially and smoothly restored with increasing baryon chemical potential $\mu_{B}$ (see the upper panel of Fig.\ref{condensates}), contrary to the first-order transition and nearly full restoration in Nambu--Jona-Lasinio model~\cite{Klevansky:1992qe}. Secondly, the vector condensate increases with $\mu_{B}$ (see the middle panel of Fig.\ref{condensates}) as the nucleon density increases, which can be easily expected from the proportionality shown in Eq.\eqref{omega01}. Thirdly, the nucleon ratio reduces (see the lower panel of Fig.\ref{condensates}) as the quark density enhances  more quickly than the nucleon density, which indicates gradual dominance of the quark degrees of freedom at larger chemical potential. Finally we have calculated the equation of state and especially the speed of sound for quarkyonic matter in this model. The results are found to be consistent with predictions of various other models as well as with experimental constraints for symmetric nuclear matter at a few times the saturation density. Our overall conclusion is that,  quantitative results from our field theoretical model of quarkyonic matter provide a satisfactory description about the properties of vacuum as well as nuclear matter up to several times the saturation density. Apart for the first-order liquid-gas transition at $\mu_{Bc}=923\,{\rm MeV}$, the results feature a continuous transition from nucleon-dominated regime to quark-dominated one along with gradual restoration of the chiral symmetry.  It will be a natural step to further explore the implications of this quarkyonic matter model for the interiors of neutron stars, such as had been done in Ref.~\cite{McLerran:2018hbz}. The results shall be reported in a future publication. 

\emph{Acknowledgments}---
The authors are grateful to Charles Horowitz and Larry McLerran for very helpful discussions. G.C. is supported by the National Natural Science Foundation of China with Grant No. 11805290 and Young Teachers Training Program of Sun Yat-sen University with Grant No. 19lgpy282. J.L. is   supported in part by the U.S. NSF Grant No. PHY-1913729 and by the U.S. DOE Office of Science, Office of Nuclear Physics, within the framework of the Beam Energy Scan Theory (BEST) Topical Collaboration.


\begin{thebibliography}{99}


\bibitem{Page:2006ud}
D.~Page and S.~Reddy,
Ann. Rev. Nucl. Part. Sci. \textbf{56}, 327-374 (2006)
doi:10.1146/annurev.nucl.56.080805.140600
[arXiv:astro-ph/0608360 [astro-ph]].

\bibitem{Lattimer:2015nhk}
J.~M.~Lattimer and M.~Prakash,
Phys. Rept. \textbf{621}, 127-164 (2016)
doi:10.1016/j.physrep.2015.12.005
[arXiv:1512.07820 [astro-ph.SR]].

\bibitem{Braun-Munzinger:2015hba}
P.~Braun-Munzinger, V.~Koch, T.~Schäfer and J.~Stachel,
Phys. Rept. \textbf{621}, 76-126 (2016)
doi:10.1016/j.physrep.2015.12.003
[arXiv:1510.00442 [nucl-th]].

\bibitem{Bzdak:2019pkr}
A.~Bzdak, S.~Esumi, V.~Koch, J.~Liao, M.~Stephanov and N.~Xu,
Phys. Rept. \textbf{853}, 1-87 (2020)
doi:10.1016/j.physrep.2020.01.005
[arXiv:1906.00936 [nucl-th]].

\bibitem{Luo:2017faz} 
  X.~Luo and N.~Xu,
  Nucl.\ Sci.\ Tech.\  {\bf 28}, no. 8, 112 (2017)
  doi:10.1007/s41365-017-0257-0
  [arXiv:1701.02105 [nucl-ex]].
  
\bibitem{Alford:1997zt} 
  M.~G.~Alford, K.~Rajagopal and F.~Wilczek,
  Phys.\ Lett.\ B {\bf 422}, 247 (1998)
  doi:10.1016/S0370-2693(98)00051-3
  [hep-ph/9711395].
  
\bibitem{Rapp:1997zu} 
  R.~Rapp, T.~Schäfer, E.~V.~Shuryak and M.~Velkovsky,
  Phys.\ Rev.\ Lett.\  {\bf 81}, 53 (1998)
  doi:10.1103/PhysRevLett.81.53
  [hep-ph/9711396].
  
\bibitem{Alford:1998mk} 
  M.~G.~Alford, K.~Rajagopal and F.~Wilczek,
  Nucl.\ Phys.\ B {\bf 537}, 443 (1999)
  doi:10.1016/S0550-3213(98)00668-3
  [hep-ph/9804403].
  
\bibitem{McLerran:2007qj} 
  L.~McLerran and R.~D.~Pisarski,
  Nucl.\ Phys.\ A {\bf 796}, 83 (2007)
  doi:10.1016/j.nuclphysa.2007.08.013
  [arXiv:0706.2191 [hep-ph]].
  
  %
\bibitem{McLerran:2008ua}
L.~McLerran, K.~Redlich and C.~Sasaki,
Nucl. Phys. A \textbf{824}, 86-100 (2009)
doi:10.1016/j.nuclphysa.2009.04.001
[arXiv:0812.3585 [hep-ph]].

\bibitem{Andronic:2009gj}
A.~Andronic, D.~Blaschke, P.~Braun-Munzinger, J.~Cleymans, K.~Fukushima, L.~McLerran, H.~Oeschler, R.~Pisarski, K.~Redlich, C.~Sasaki, H.~Satz and J.~Stachel,
Nucl. Phys. A \textbf{837}, 65-86 (2010)
doi:10.1016/j.nuclphysa.2010.02.005
[arXiv:0911.4806 [hep-ph]].

\bibitem{Kojo:2009ha}
T.~Kojo, Y.~Hidaka, L.~McLerran and R.~D.~Pisarski,
Nucl. Phys. A \textbf{843}, 37-58 (2010)
doi:10.1016/j.nuclphysa.2010.05.053
[arXiv:0912.3800 [hep-ph]].

\bibitem{Cao:2016ats} 
G.~Cao, L.~He and X.~G.~Huang,
Chin.\ Phys.\ C {\bf 41}, no. 5, 051001 (2017)
doi:10.1088/1674-1137/41/5/051001
[arXiv:1610.06438 [nucl-th]].

\bibitem{Steinheimer:2011ea} 
J.~Steinheimer, S.~Schramm and H.~Stocker,
Phys.\ Rev.\ C {\bf 84}, 045208 (2011)
doi:10.1103/PhysRevC.84.045208
[arXiv:1108.2596 [hep-ph]].

\bibitem{McLerran:2018hbz} 
L.~McLerran and S.~Reddy,
Phys.\ Rev.\ Lett.\  {\bf 122}, no. 12, 122701 (2019)
doi:10.1103/PhysRevLett.122.122701
[arXiv:1811.12503 [nucl-th]].

\bibitem{Fukushima:2015bda}
K.~Fukushima and T.~Kojo,
Astrophys. J. \textbf{817}, no.2, 180 (2016)
doi:10.3847/0004-637X/817/2/180
[arXiv:1509.00356 [nucl-th]].


\bibitem{Jeong:2019lhv} 
K.~S.~Jeong, L.~McLerran and S.~Sen,
arXiv:1908.04799 [nucl-th].

\bibitem{Sen:2020peq}
S.~Sen and N.~C.~Warrington,
[arXiv:2002.11133 [nucl-th]].

\bibitem{Duarte:2020xsp}
D.~C.~Duarte, S.~Hernandez-Ortiz and K.~S.~Jeong,
[arXiv:2003.02362 [nucl-th]].

\bibitem{Zhao:2020dvu}
T.~Zhao and J.~M.~Lattimer,
[arXiv:2004.08293 [astro-ph.HE]].

\bibitem{Xia:2018cpy}
C.~J.~Xia, S.~S.~Xue and S.~G.~Zhou,
JPS Conf. Proc. \textbf{20}, 011010 (2018)
doi:10.7566/JPSCP.20.011010


\bibitem{Schaefer:2006ds} 
B.~J.~Schaefer and J.~Wambach,
Phys.\ Rev.\ D {\bf 75}, 085015 (2007)
doi:10.1103/PhysRevD.75.085015
[hep-ph/0603256].

\bibitem{Walecka}
J.D. Walecka, Ann. of Phys. 83, 491 (1974).

\bibitem{Klevansky:1992qe}
S.~P.~Klevansky,
"The Nambu-Jona-Lasinio model of quantum chromodynamics,"
Rev.\ Mod.\ Phys.\  {\bf 64}, 649 (1992).

\bibitem{Hofstadter} R. Hofstadter, Rev. Mod. Phys., 28:214 (1956).

\bibitem{Green1953} A. E. S. Green and D. F. Edwards, Phys. Rev., 91:46 (1953).

\bibitem{Green1954} A. E. S. Green, Phys. Rev., 95:1006 (1954). 

\bibitem{Shlomo2006}
S. Shlomo, V. M. Kolomietz, G. Col`o, Eur. Phys. J. A 30, 23-30 (2006).

\bibitem{Baym:2017whm} 
  G.~Baym, T.~Hatsuda, T.~Kojo, P.~D.~Powell, Y.~Song and T.~Takatsuka,
  Rept.\ Prog.\ Phys.\  {\bf 81}, no. 5, 056902 (2018)
  doi:10.1088/1361-6633/aaae14
  [arXiv:1707.04966 [astro-ph.HE]].
  
\bibitem{Leonhardt:2019fua} 
M.~Leonhardt, M.~Pospiech, B.~Schallmo, J.~Braun, C.~Drischler, K.~Hebeler and A.~Schwenk,
arXiv:1907.05814 [nucl-th].

 \bibitem{Danielewicz2002} 
P. Danielewicz, R. Lacey and W.G. Lynch, Science 298 (2002) 1592-1596.

\end{thebibliography}
\end{document}